\pdfoutput=1

\documentclass{raa}

\usepackage{mathptmx}
\usepackage{ae,aecompl}

\usepackage{graphicx}	
\usepackage{amsmath}	
\usepackage{amssymb}	
\usepackage{pifont}
\usepackage{txfonts}
\usepackage{url}
\usepackage{subfigure}
\usepackage{booktabs}
\usepackage{longtable}
\usepackage{lscape}
\usepackage{multirow}
\usepackage{color}
\usepackage{textcomp}
\usepackage{natbib}
\usepackage[colorlinks,linkcolor=red,anchorcolor=green,citecolor=blue]{hyperref}
\usepackage{threeparttable}
\usepackage{enumerate}

\begin{document}

\title{Morphology Study for GeV Emission of the Nearby Supernova Remnant G332.5-5.6}


   \volnopage{Vol.0 (2024) No.0, 000--000}      
   \setcounter{page}{1}          
   \author{Ming-Hong Luo\inst{1},
      Qing-Wen Tang\inst{1}$^{,\star}$,
      and Xiu-Rong Mo\inst{1}
   }

   \institute{
    \inst{1}Department of Physics, School of Physics and Materials Science, Nanchang University, Nanchang 330031, China\\}
     \email{qwtang@ncu.edu.cn}

\abstract{Spatial template is important to study the nearby supernova remnant (SNR). For SNR G332.5-5.6, we report a gaussian disk with radius of about 1.06\textdegree \ to be a potential good spatial model in the $\gamma$-ray band. Employing this new gaussian disk, its GeV lightcurve shows a significant variability of about seven sigma. The $\gamma$-ray observations of this SNR could be explained well either by a leptonic model or a hadronic model, in which a flat spectrum for the ejected electrons/protons.
 }

   \authorrunning{Luo et al.}                     
   \titlerunning{Morphology of GeV Emission from SNR G332.5-5.6}       

   \maketitle


\section{Introduction}    
\label{sec:Introduction}

The energy released by supernova events is of great importance for our understanding of the physics of the interstellar medium (ISM). Depending upon the environments, the supernova remnants (SNRs) can display a vast range of shapes~\citep{1996A&AS..118..329W}. The released energy come into the electrons and protons, which interact with the interstellar radiation or protons, and finally emit the observable electromagnetic signals up to high-energy bands ($>$ 100 MeV)~\citep{2016ApJS..224....8A}.

In the Anglo Australian Observatory (AAO)/United Kingdom Schmidt Telescope (UKST) H$\alpha$ survey, a new Galactic supernova remnant (SNR)
was uncovered~\citep{2005MNRAS.362..689P}. This source with an unusual morphology, dubbed the "paperclip", was discovered from the original H$\alpha$ survey films, which was further confirmed as SNR G332.5-5.6~\citep{2007MNRAS.381..377S}. 
SNR G332.5-5.6 shows three patches of filamentary emission with a total size of about 30 arcmin at the radio band~\citep{2007MNRAS.375...92R}.
It has a extended X-rays morphology in the center region, which has a good correlation with radio emission detected at different frequencies~\citep{2015AA...583A..84S}. 
SNR G332.5-5.6 is found to be a distance about 3.4 kpc with an age of 7-9 kyr~\citep{2015MNRAS.452.3470Z}. 
For its $\gamma$-ray observations, such as in the \textit{Fermi} High-Latitude Extended Sources (FHES) catalog, SNR G332.5-5.6 is the potential association of FHES J1642.1-5428, which has a uniform disk with radius of about 0.57\textdegree~\citep{2018ApJS..237...32A}. In an incremental version of the fourth full catalog of Fermi-LAT sources (4FGL-DR3), SNR G332.5-5.6 associates with an extended source of 4FGL J1642.1-5428e, which shares a disk radius of about 0.70\textdegree~\citep{2022ApJS..260...53A}.

Morphology in the $\gamma$-ray band is very different from that in the low-energy bands, such as radio or X-ray bands. Since the FHES catalog employed only 8 years of observations by \textit{Fermi} Large Area Telescope (Fermi/LAT), the spatial model of SNR G332.5-5.6 could be studied in deep employing more gamma-ray observations. We thus could performed morphology study in the high-energy gamma-ray bands, such as 14.3-year Fermi-LAT observations. The paper is organized as follows. In \S~\ref{sec:data}, data analysis is presented. In \S~\ref{sec:model}, the physical origin of GeV emission of SNR G332.5-5.6 is explored. Summary and conclusions are presented in \S~\ref{sec:conclusion}. 

\section{Fermi/LAT Data Analysis}
\label{sec:data}

\subsection{Data Selection}
 $FermiPy$ version 1.2 package~\footnote{\url{https://fermipy.readthedocs.io/en/v1.2/}} is used to analyze the Fermi/LAT data, which is a python package that facilitates analysis of the Fermi-LAT data with the latest Fermi Science Tools version 2.2.0~\footnote{\url{https://github.com/fermi-lat/Fermitools-conda/wiki/Installation-Instructions}}~\citep{2017ICRC...35..824W}. Pass 8 events are selected in the region of interest (ROI) of 15\textdegree around the position of SNR
G332.5-5.6, that is R.A., decl. = 250.538\textdegree, -54.477\textdegree~\citep{2017ApJS..232...18A}. We chose the photons in the energy range between 100 MeV and 1 TeV within 2008 August and 2022 December. The events were selected with event class of 128
and event type of 3. Events with zenith angles larger than 90\textdegree \ are excluded to avoid
the contamination from the Earth's limb. We use the standard data quality selection criteria {(DATA\_QUAL$>$0)\&\&(LAT\_CONFIG==1)}. We accepted a pixel size of 0.02\textdegree \ spatially and 8 logarithmic energybins per decade when performing the energy selection. {P8R3\_SOURCE\_V3} instrument response function is used in our data analysis.

\subsection{Basic Model}
\begin{figure}[h]
  \centering
  \includegraphics[width=0.7\linewidth]{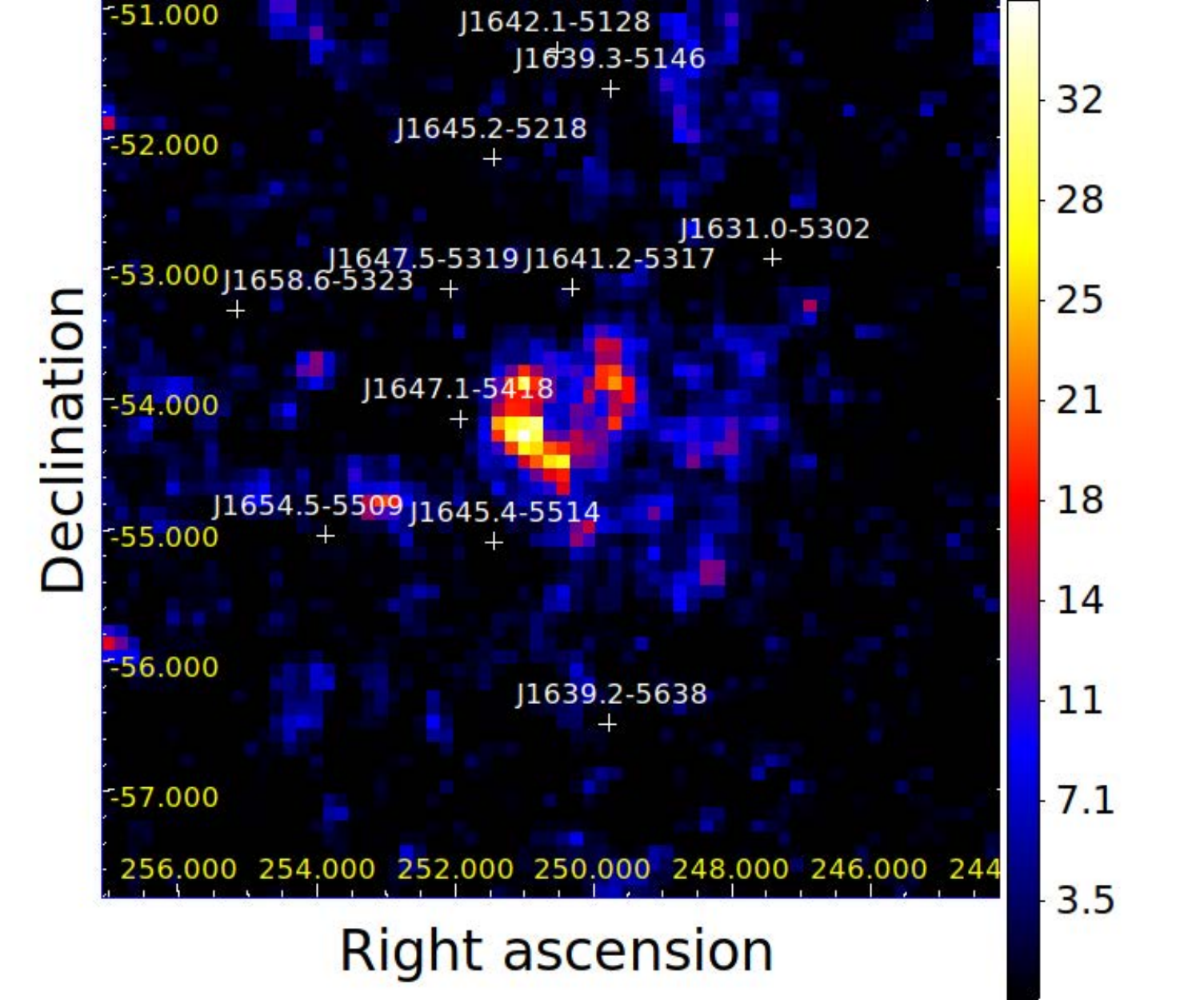}
  \caption{TS map  (7\textdegree$\times$7\textdegree) of the NONE model with pixel size of 0.1$^{\circ}$ between 0.1-1000 GeV centering at SNR G332.5-5.6. Sources marked by white cross are 11 background point sources. \label{fig:ts}}
\end{figure}

A basic model is built by adding the background gamma-ray sources. 4FGL-DR3 is employed, in which we include all sources that 25\textdegree \ within SNR G332.5-5.6 center, as well as the diffuse galactic interstellar emission ({gll\_iem\_v07.fits}) and the isotropic emission ({iso\_P8R3\_SOURCE\_V3\_V1.txt}). The spectral parameters of the diffuse galactic interstellar emission, the isotropic emission and all sources that 3 degrees within SNR G332.5-5.6 center are allowed to be free. We found there are 11 point sources as shown in Figure~\ref{fig:ts}, whose spectral parameters are allowed to be free. Lastly, we removed the catalog source of 4FGL J1642.1-5428e in the ROI center and then built a basic model (hereafter the model with the spatial template of NONE).

\subsection{Spatial Analysis}
We evaluate the significance of all test spatial models, which is quantified through the likelihood ratio test with two statistical method.
One is the test statistic (TS) to compute the significance for a test model over the NONE spatial model, which is defined as
	\begin{equation}
		{\rm TS}=2\times (ln\mathcal{L}_{\rm test}- ln\mathcal{L}_{\rm NONE}),
	\end{equation}
Where $\mathcal{L}_{\rm test}$ and $\mathcal{L}_{\rm NONE}$ are the maximum likelihood values of the
test model and of the basic NONE model respectively~\citep{1996ApJ...461..396M}. The likelihood value depends on the model and the observations. For the Fermi-LAT observations, the photon numbers in the ROI are 9065267, 1691719, 56799, 2052 in the energy bands of 0.1-1 GeV, 1-10 GeV, 10-100 GeV and 100-1000 GeV respectively. The other one is the Akaike information criterion (AIC) test to find out a better model between two nest models or unnest models~\citep{1974ITAC...19..716A,2018A&A...617A..78T,2021ApJ...922..255T,2024arXiv240211880G}, which is defined as
\begin{equation}
    {\rm AIC}=2k-2ln\mathcal{L}
\end{equation}
Where $k$ is the number of degrees of freedom (d.o.f) and $\mathcal{L}$ is the maximum likelihood value of a model. Usually, a model with a smaller AIC is the better one among several models. In this case, we employed the difference in AIC ($\Delta$AIC) between two models to claim the prefer model, which could be calculated by 
\begin{equation}
\Delta {\rm AIC} = 2\Delta k - 2\Delta ln\mathcal{L}.
\end{equation}

\begin{figure}[h]
  \centering
  \includegraphics[width=\linewidth]{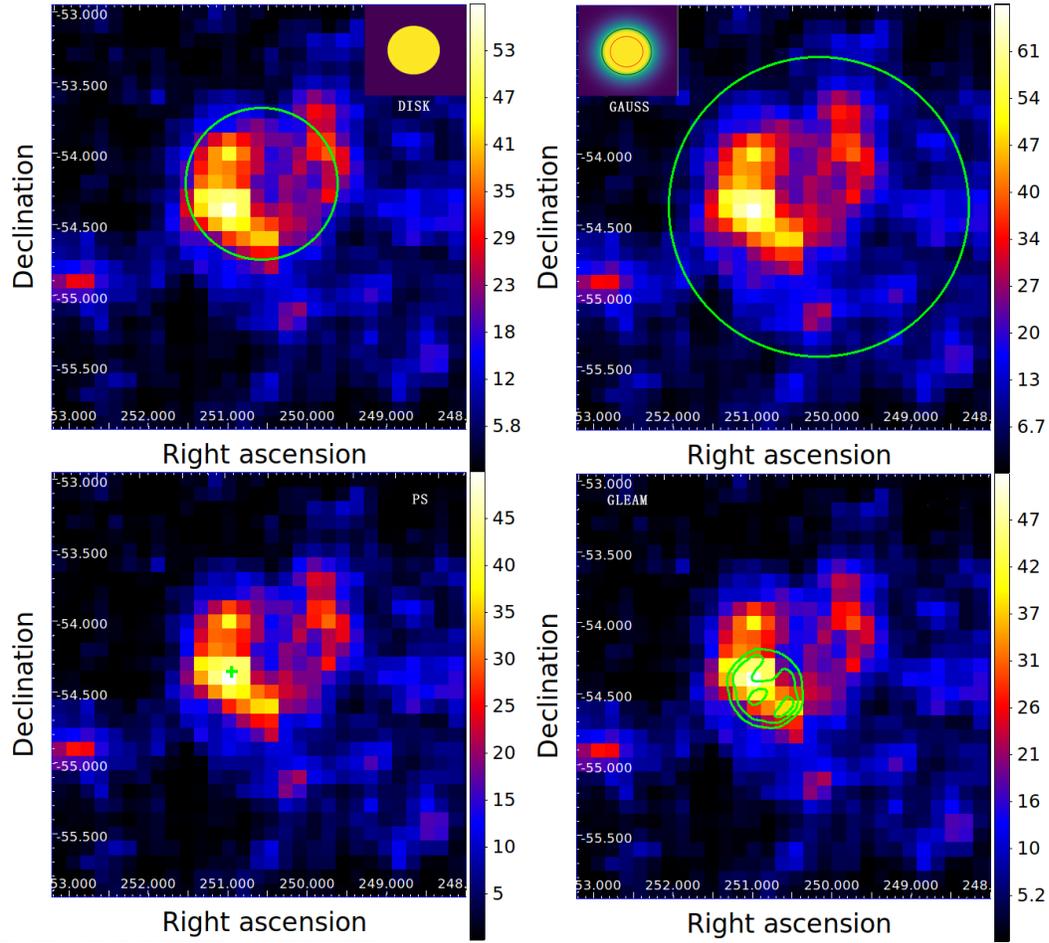}
  \caption{TS maps  (5\textdegree$\times$5\textdegree) after removing SNR G332.5-5.6 in four spatial templates. Top left: DISK template with radius of $\sim$0.54$^{\circ}$. Top right: GAUSS template with a radius of $\sim$1.06$^{\circ}$. Bottom left: PS template. Bottom right: GLEAM template. Green circle represent the corresponding radius, green cross is the point source and green contour is from the radio observations. \label{fig:spatial}}
\end{figure}

Four spatial templates are tested with a single power-law (PL) spectral model, as shown in Table~\ref{tab:morphology}. TS maps excluding SNR G332.5-5.6 in each model is plotted in Figure~\ref{fig:spatial}.

\begin{enumerate}[I.]
  \item PS model. It shares the spatial template of a point source, which has four additional free parameters (two for the free position and two PL spectral parameters). The resultant maximum TS is 35.16 and the best-fit position is about R.A., decl. = 250.925\textdegree, -54.388\textdegree.
  \item GLEAM model. It is derived from radio observations of Galactic and Extragalactic Allsky MWA Survey between 170 MHz and 231 MHz is employed~\citep{2015PASA...32...25W}, see Figure~\ref{fig:GLEAM}, which has two additional free parameters(two PL spectral parameters). The resultant maximum  TS is 71.36.
  \item DISK model. It is a uniform disk, which has five additional free parameters (One for free radius, two for the free centered position and two PL spectral parameters). We scanned the disk radius from 0.01\textdegree \ to 1.20\textdegree \ using the tool of source finding. We found a best-fit radius of 0.536\textdegree \ with the maximum TS of 223.34. The best-fit centered position is about R.A., decl. = 250.564\textdegree, -54.245\textdegree.
  \item GAUSS model. It is a two-dimensional gaussian disk, which also has five additional free parameters. The scanning method is same as that in the DISK model. The best-fit 68\% containment radius is found at about 1.061\textdegree \ with the maximum TS of 249.94. The best-fit centered position is about R.A., decl. = 250.173\textdegree, -54.401\textdegree.
\end{enumerate}

Results of AIC test are shown in Table~\ref{tab:morphology}. As seen, GAUSS model is confirmed to be the best representation of the
data, with a minimum $\Delta$AIC of -239.94 respect to the NONE model. Combing that GAUSS also has the largest TS value among four test spatial models, GAUSS model is the preferred model in our spatial analysis. In the latest Fermi-LAT catalog (4FGL-DR4), there are three SNRs with spatial template of GAUSS, such as SNR G51.3+0.1, SNR G150.3+4.5 and SNR G292.2-0.5~\footnote{\url{https://fermi.gsfc.nasa.gov/ssc/data/access/lat/14yr_catalog/}}. Therefore GAUSS spatial template is selected for the SNR G332.5-5.6 to produce its lightcurve (LC) and the spectral energy distribution (SED) in the following analyses. 

\subsection{Temporal Analysis}

\begin{figure}[h]
  \centering
  \includegraphics[width=\linewidth]{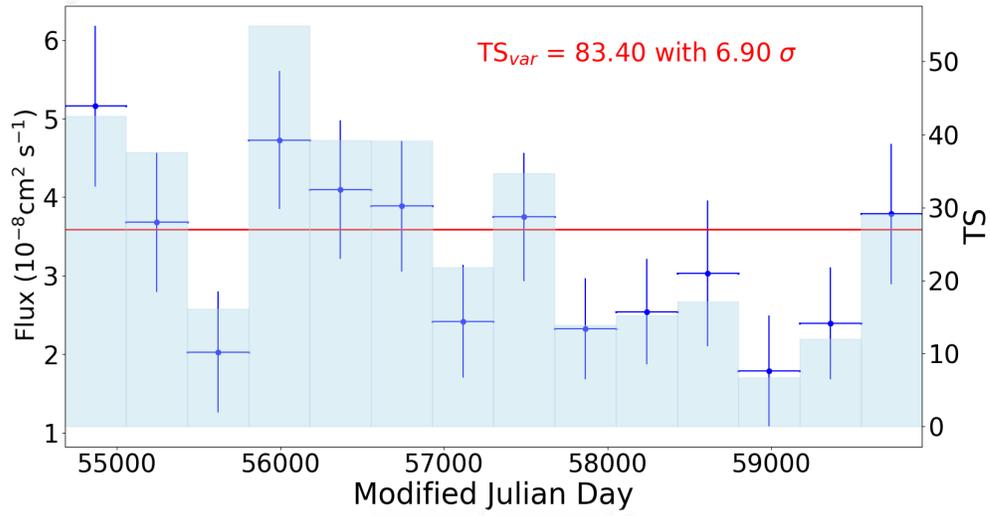}
  \caption{Lightcurve of SNR G332.5-5.6 with 14 time bins in the energy band of 0.1-1000 GeV. Data points are plotted with errors at confidence levels 95$\%$. The light-blue shadows are the TS values. The red line is the constant flux employing the full time observation of 14.3 years. \label{fig:lcs}}
\end{figure}

To test whether GeV emission of SNR G332.5-5.6 is variable temporally, we construct its lightcurve in GAUSS spatial model with a PL spectral model. We separate Fermi-LAT observations into 14 time intervals in the energy range from 0.1 GeV to 1 TeV. Energy flux in each time interval is calculated, which is plotted in Figure~\ref{fig:lcs}. The variability index (${\rm TS}_{var}$) is defined as
\begin{equation}
		{\rm TS}_{var}= -2\sum_{i}^{N}  ln\frac{\mathcal{L}_{i}(F_{const}) }{\mathcal{L}_{i}(F_{i})}
		\label{TSvar}
	\end{equation}
where $\mathcal{L}_{i}$ is the value of the likelihood corresponding to the $i$-th bin, $F_i$ is the best-fit flux for bin $i$, and $F_{const}$ is the best-fit flux for the full time assuming a constant flux ~\citep{2012ApJS..199...31N,2023RAA....23b5007M}. The resultant ${\rm TS}_{var}$ is 83.40 with 6.9 $\sigma$ for the LC of fourteen time bins, which suggests that there has a significant variability for the photon fluxes of SNR G332.5-5.6.

\subsection{Spectral Analysis}

\begin{figure}[h]
  \centering
  \includegraphics[width=\linewidth]{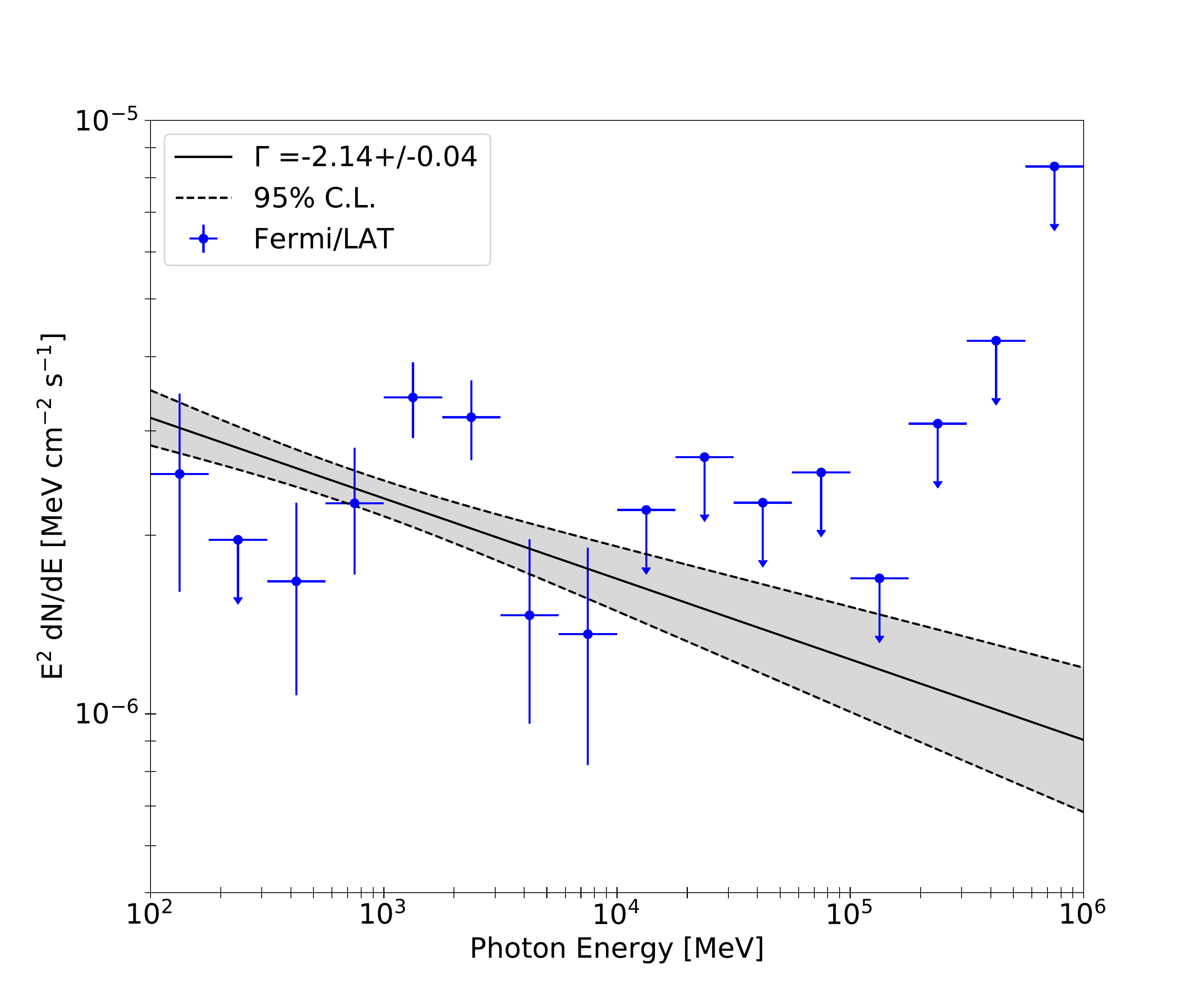}
  \caption{Spectral energy distribution of SNR G332.5-5.6. Data points are plotted with 95$\%$ errors when TS $>$ 5, while the upper limits are plotted at the 95$\%$ confidence levels when TS $<$ 5. Black solid line represents the best fit while the black dotted lines are the errors at 95$\%$ confidence levels. \label{fig:sed}}
\end{figure}

Gamma-ray emission of the analyzed source is represented by a PL spectral model in the 4FGL-DR3 catalog. To fully investigate the $\gamma$-ray properties of this source, we considered other types of spectral models, such as broken PL (BPL), log-parabola (LP) and PLSuperExpCutoff (PLEC), which however do not resulted in any significant likelihood improvement. We thus employed a PL spectral model to fit the GeV emission of SNR G332.5-5.6, e.g., $\frac{dN}{dE}\propto E^{-\Gamma_{\rm ph}}$.  The resultant photon index ($\Gamma_{\rm ph}$) is 2.14$\pm$0.04. The spectral energy distribution is plotted in Figure~\ref{fig:sed}. 

\section{Physical Model}
\label{sec:model}

We explore the physical origins utilizing Fermi/LAT observations of SNR G332.5-5.6. $Naima$ package~\footnote{\url{https://naima.readthedocs.io/en/latest/index.html}} is employed to find the best-fit parameters in two physical models, such as a leptonic model and a hadronic model~\citep{2015ICRC...34..922Z}. 

\subsection{Leptonic Model}
\begin{figure}[h]
  \centering
  \includegraphics[width=\linewidth]{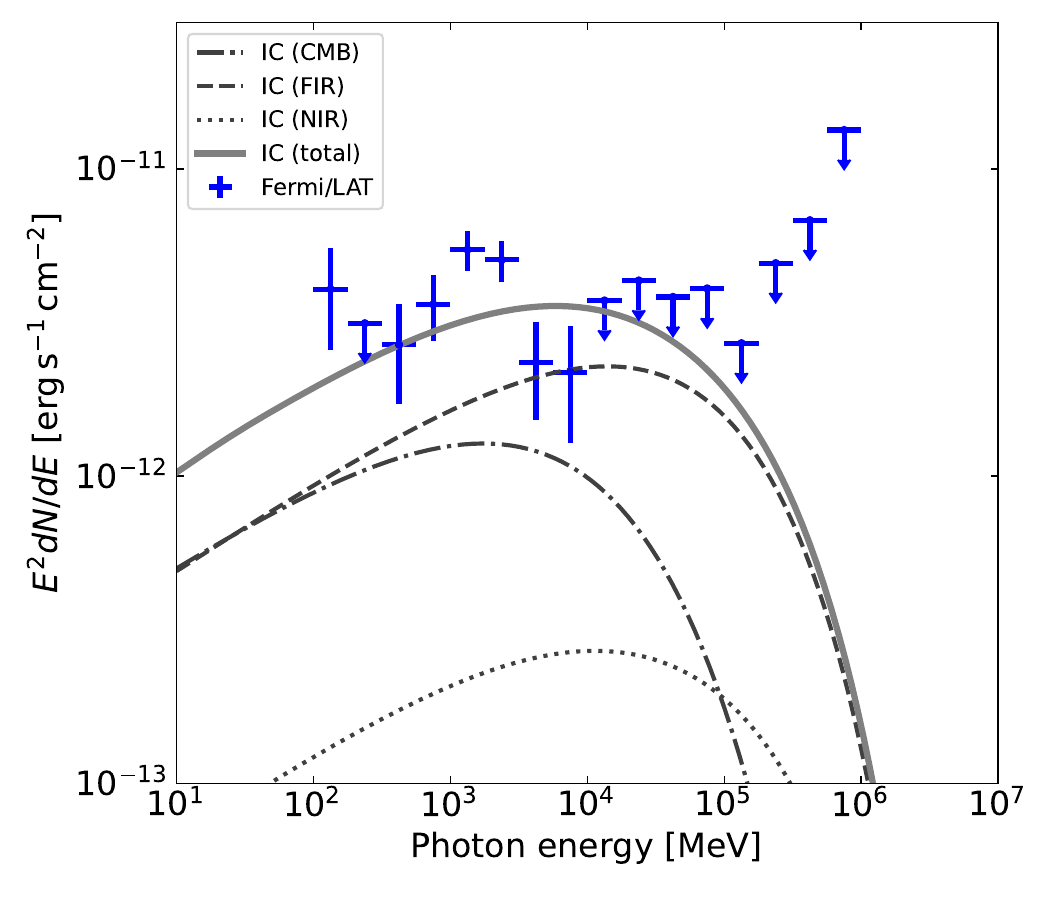}
  \caption{Theoretical expectation of the leptonic model for GeV SED of SNR G332.5-5.6. Total IC components contribute to gamma-ray emission detected by Fermi-LAT. Here, the electron injection spectral index is $\alpha_e$ = 2.01. \label{fig:physicalmodel1}}
\end{figure}

We considered a leptonic model to fit observed GeV SED of SNR G332.5-5.6, e.g., the $\gamma$-ray emission is generated through inverse Compton (IC) scattering of soft photons by relativistic electrons~\citep{2006ApJ...647..692K,2010A&A...523A...2M,2018Ap&SS.363...25T}. The ejected electronic distribution is assumed to follow an exponential-cutoff power-law function (PLEC),
\begin{equation}
	F(E)=A(\frac{E}{E_{0} })^{-\alpha_e}\exp(-\frac{E}{E_{\rm cutoff} })
	\label{PLEC}
\end{equation}
Where $A$ is the normalization, $E_{0}$ is the pivot energy fixed at 1~TeV , $\alpha_e$ is the electron injection spectral index, and $E_{\rm cutoff}$ stands for the cutoff energy. For IC, seed photons include (1) the cosmic microwave background (CMB), (2) the far-infrared dust emission (FIR) with the temperature of 30 K and energy density of 0.5 eV/cm$^{3}$, and (3) the near-infrared stellar emission (NIR) with the temperature of 5000 K and energy density of 1.0 eV/cm$^{3}$. The best-fit results are shown in Figure~\ref{fig:physicalmodel1} and in Table~\ref{tab:physicalmodels}. A electron injection spectral index of $\alpha_e$ = 2.01 is required to produce the gamma-ray emission above 100~MeV. The total energy of the injection electrons is about 3.08 $\times 10^{48}$ ergs.

\subsection{Hadronic Model}
\begin{figure}[h]
  \centering
  \includegraphics[width=\linewidth]{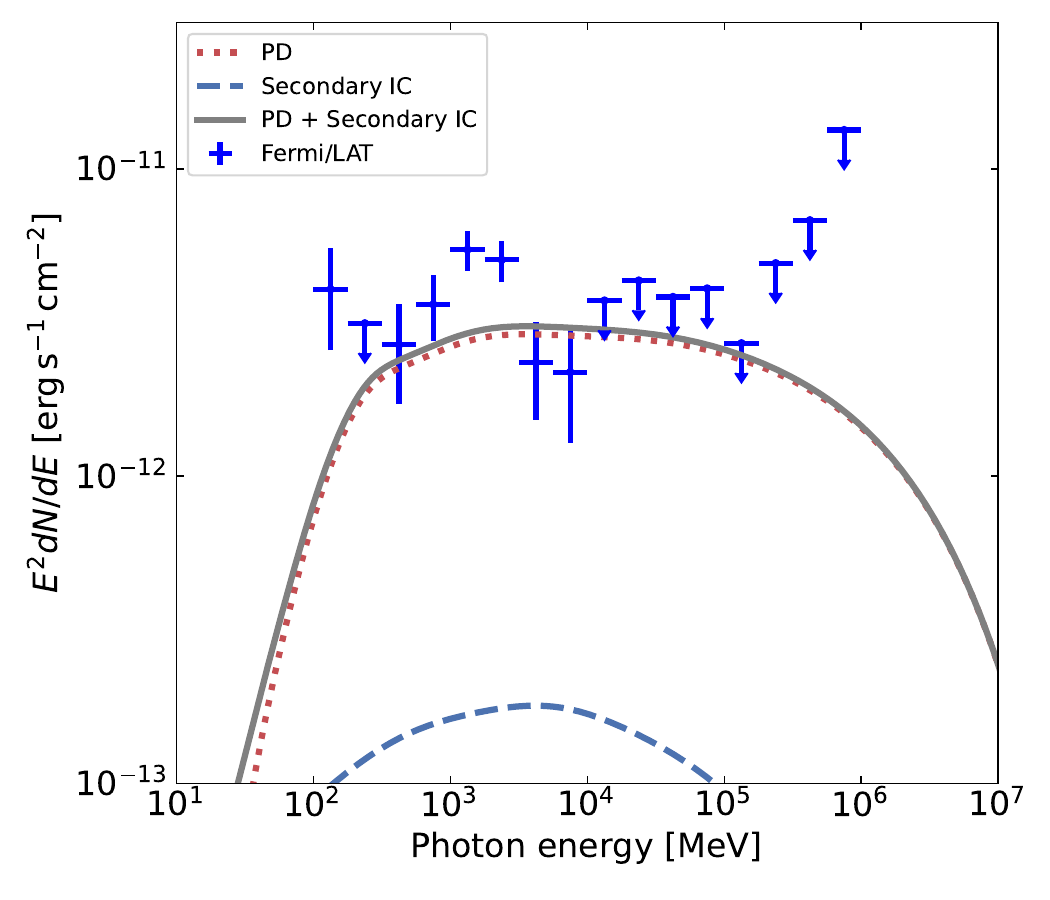}
  \caption{Hadronic-model fitting for GeV SED of SNR G332.5-5.6. PD and secondary IC components contribute to gamma-ray emission detected by Fermi-LAT. Here the proton spectral index is $\alpha_p$ = 2.03 and the gas density is $n_{\rm gas} = 1.00\ {\rm cm}^{-3}$. \label{fig:physicalmodel2}}
\end{figure}

We considered a one-zone hadronic model~\citep{2006ApJ...647..692K,2017ApJ...843...42T,2023RAA....23b5007M}, in which the main gamma-ray production for relativistic protons are $p-p$ interactions followed by pion decay (PD). In addition to gamma-rays, secondary electrons and positrons are produced in the PD process, which result in an inverse Compton emission (Secondary IC). Thus, the PD component plus a secondary IC component contributes for the Fermi-LAT observations. The distribution of the ejected protons is also assumed to follow a PLEC function,
\begin{equation}
	F(E)=A(\frac{E}{E_{0} })^{-\alpha_p}\exp(-\frac{E}{E_{\rm cutoff} })
	\label{PLEC2}
\end{equation}
Where $A$ is the normalization, $E_{0}$ is the pivot energy fixed at 10~TeV , $\alpha_p$ is the spectral index of the injected protons, and $E_{\rm cutoff}$ represents the cutoff energy. The best-fit results are shown in Figure~\ref{fig:physicalmodel2} and in Table~\ref{tab:physicalmodels}. We performed the fit and derived the best-fit parameters: $\alpha_p$ = 2.03, $E_{\rm cutoff}$ = 29.82 TeV and the gas density of $n_{\rm gas} = 1.00\ {\rm cm}^{-3}$. The total energy of the injection protons is about 1.32 $\times 10^{50}$ ergs. 

In summary, both leptonic model and hadronic model can fit GeV SED of SNR G332.5-5.6 well, which share a flat spectrum for the ejected electrons or protons, e.g., spectral indices are close to -2.00. 

\section{Summary and Conclusion}
\label{sec:conclusion}

In this work, we present a GeV morphology analysis of SNR G332.5-5.6 employing the Fermi/LAT observations from 2008 August to 2022 December. We found that a gaussian-disk spatial template with radius of 1.06\textdegree \ is the preferred one with respect to others, e.g., a point source, a uniform disk and an extended template from the radio observations. Employing the new gaussian disk, the GeV lightcurve for SNR G332.5-5.6 shows a significant variability. The gamma-ray observations could be explained well either by a leptonic model or a hadronic model, in which the flat spectrum for the ejected electrons or protons is required. More future observations in other bands could shed light on the nature of broadband emission from SNR G332.5-5.6.

\section*{Acknowledgements}
\addcontentsline{toc}{section}{Acknowledgements}

We thank the anonymous referee for his/her comments and suggestions that improved the manuscript significantly. We also thank Ruo-Yu Liu for helpful discussions. This work is supported by the National Natural Science Foundation of China under grant 12065017, and the Jiangxi Provincial Natural Science Foundation of China under grants Nos. 20224ACB211001 and 20212BAB201029.

\newpage

\begin{table}[!ht]
    \caption{Spatial Analysis Results between 100 MeV and 1 TeV}
    \label{tab:morphology}
    \centering
    \begin{threeparttable}        
      \begin{tabular}{cccccccc}
    \toprule
    \ Spatial Template\tnote{a} & R.A.  &Decl. &$R_{68}$ &$\Delta$ k &$-{\rm log} ({\rm Likelihood})$  & $\Delta$AIC \\
    & (degree)   & (degree) &(degree) &  &  &  &   \\
    \midrule
    \  NONE  & - & - & -        & 0 &-16292446.11 &  0 \\
    \ PS     & 250.925$\pm$0.063 &-54.388$\pm$0.065 &-      & 4 &-16292463.69 &  -27.16 \\
    \ GLEAM  & - &- &-          & 2 &-16292481.79 &  -67.36  \\
    \ DISK   & 250.564$\pm$0.065 &-54.245$\pm$0.075    &0.536$\pm$0.037 & 5 &-16292557.78 &  -213.34 \\
    \ GAUSS  & 250.173$\pm$0.131 &-54.401$\pm$0.117    &1.061$\pm$0.153 & 5 &-16292571.08 &  -239.94 \\
    \bottomrule
      \end{tabular}
         \begin{tablenotes} 
        \footnotesize          
        \item[a] Spatial template for SNR G332.5-5.6, NONE represents null source, GLEAM is a extended map by radio observation, GAUSS is the gaussian disk, PS is a single point source, DISK is the uniform disk. 
      \end{tablenotes}           
    \end{threeparttable}      
  \end{table}

\begin{table}[!ht]
    \caption{Best-fit physical parameters in leptonic model and hadronic model}
     \label{tab:physicalmodels}
    \centering
    \begin{threeparttable}        
      \begin{tabular}{*5{c}}
    \toprule
    \ Physical Model & log$_{10}$($A$) &  $\alpha_e$/$\alpha_p$ &  $E_{\rm cutoff}$ &  \\
       \  & (${\rm GeV}^{-1}$) &  &  ({\rm TeV})  & ($10^{48}$ erg)\\
    \midrule
    \ IC              & 41.06  &2.01  & 1.05    \\
    \ PD+Secondary IC & 52.88  &2.03  & 29.82   \\
    \bottomrule
      \end{tabular}        
    \end{threeparttable}      
  \end{table}

\appendix

\section{Radio Observations and the GLEAM Spatial Template}
\begin{figure}[h]
  \centering
  \includegraphics[width=\linewidth]{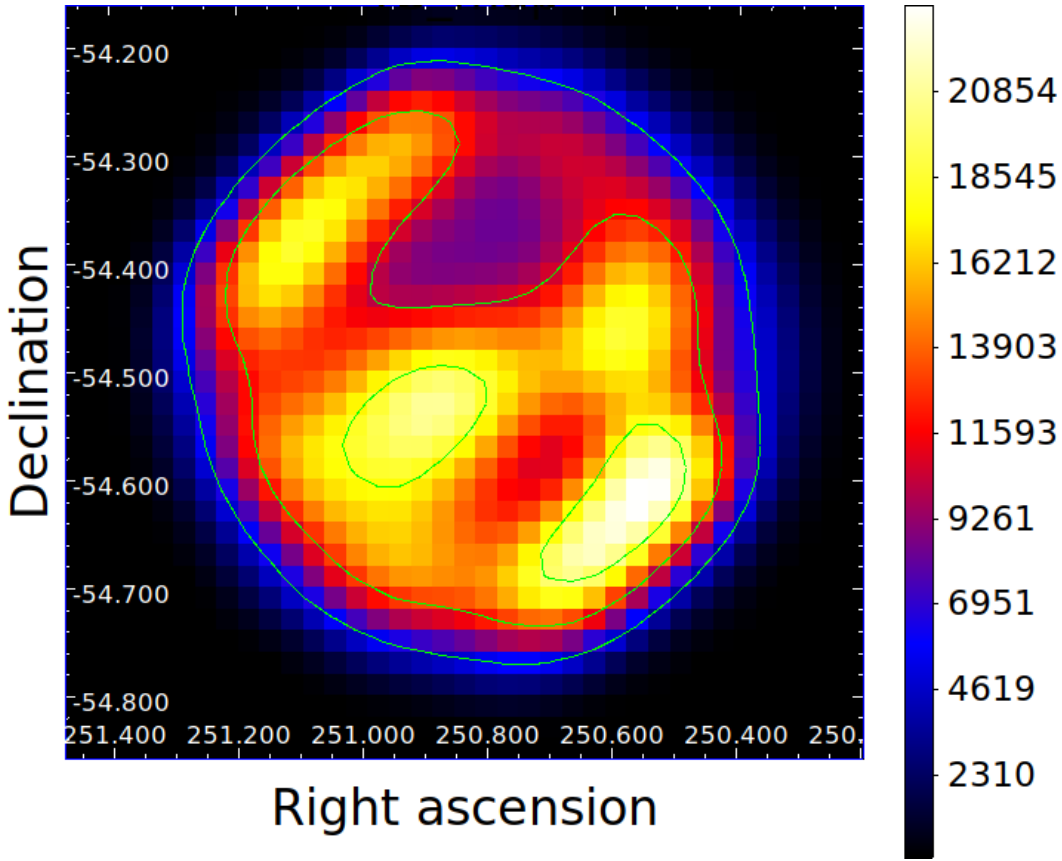}
  \caption{GLEAM spatial template (1\textdegree$\times$1\textdegree) from the ROI center, which is derived from radio observations of Galactic and Extragalactic Allsky MWA Survey bwtween 170 Mhz and 231 Mhz~\citep{2015PASA...32...25W}. Gaussian smoothing is used with bin size of 0.01 degrees. 
  \label{fig:GLEAM}}
\end{figure}

GLEAM spatial template is derived from radio observations of Galactic and Extragalactic Allsky MWA Survey bwtween 170 Mhz and 231 Mhz~\citep{2015PASA...32...25W}, the gaussian smoothing is performed with bin size of 0.01 degrees, which can be found in Figure~\ref{fig:GLEAM}.

\end{document}